\documentclass[%
 reprint,
superscriptaddress,
 amsmath,amssymb,
 aps,
longbibliography,
prl
]{revtex4-2}

\usepackage{xparse}

\usepackage{appendix}
\usepackage{graphicx}% Include figure files
\graphicspath{ {./Figures/} }

\usepackage{dcolumn}% Align table columns on decimal point
\usepackage{bm}% bold math
\usepackage{hyperref}% add hypertext capabilities
\usepackage{xcolor} % added by the author

\usepackage{comment}
\usepackage{dsfont}

\newcommand{\ket}[1]{\ensuremath{\left| #1 \right>}}
\newcommand{\bra}[1]{\ensuremath{\left< #1 \right|}}
\newcommand{\braket}[2]{\ensuremath{\left< #1 \ \vphantom{#2} \right| 
\left. #2 \vphantom{#1} \right>}}

\usepackage{slashed}
\renewcommand{\vec}[1]{\mathbf{#1}}

\usepackage{bm}
\newcommand{\vect}[1]{\boldsymbol{\mathbf{#1}}}

\newcommand{\updownarrows}{{\uparrow\!\downarrow}}

\newcommand{\dtri}{{\partial\triangledown}}
\newcommand{\tri}{\triangledown}

\usepackage{bbold}

\begin{document}
\title{A Generic Topological Criterion for Flat Bands in Two Dimensions}
% or
% Anomaly and Flat Bands in Bilayer Graphene

    \author{Alireza Parhizkar}
\affiliation{Joint Quantum Institute, University of Maryland, College Park, MD 20742, USA}

    \author{Victor Galitski}
\affiliation{Joint Quantum Institute, University of Maryland, College Park, MD 20742, USA}
\affiliation{Center for Computational Quantum Physics, The Flatiron Institute, New York, NY 10010, United States}

%\date{\today}

\begin{abstract}
  We show that the continuum limit of moir\'e graphene is described by a $(2+1)$-dimensional field theory of Dirac fermions coupled to two classical vector fields: a periodic gauge and spin field. We further show that the existence of a flat band  implies an effective dimensional reduction, where the time dimension is ``removed.''  The resulting two-dimensional Euclidean theory contains the chiral anomaly. The associated Atiyah-Singer index theorem provides a self-consistency condition for flat bands. In the Abelian limit, where the spin field is disregarded, we reproduce a periodic series of quantized magic angles known to exist in twisted bilayer graphene in the chiral limit. However, the results are not exact. If the Abelian field has zero total flux, perfectly flat bands can not exist, because of the leakage of edge states into neighboring triangular patches with opposite field orientations. We demonstrate that the non-Abelian spin component can correct this and completely flatten the bands via an effective renormalization of the Abelian component into a configuration with a non-zero total flux.  We present the \textit{Abelianization} of the theory where the Abelianized flat band can be mapped to that of the lowest Landau level. We show that the Abelianization corrects the values of the magic angles consistent with numerical results. We also use this criterion to prove that an external magnetic field splits the series into pairs of magnetic field-dependent magic angles associated with flat moir\'e-Landau bands. The topological criterion and the Abelianization procedure provide a generic practical method for finding flat bands in a variety of material systems including but not limited to moir\'e bilayers. 
\end{abstract}

\date{\today}

\maketitle
Moir\'e phenomena in twisted bilayer graphene~\cite{GrapheneWithTwist,MacDonald,NonAbelianGauge,Origin,Exp2018unconventional,Exp2018correlated,GrapheneWithTwist2020,balents,ElectronicSpectrum,BandStructure} and other systems~\cite{EmEnSc,MoireGravity} have been an active topic of research in recent years. Of particular interest for the former are flat bands, where localized electrons exhibit a variety of strongly-correlated phases. In this Letter  we draw a universal picture for the occurrence of the flat bands through emergent  gauge fields and the anomaly.

\begin{figure}
\includegraphics[width=\linewidth]{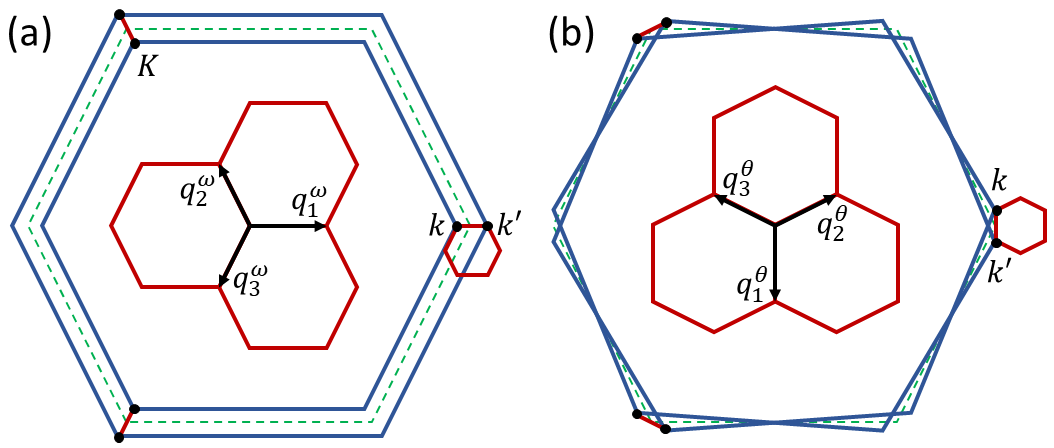}
\centering
\caption{Each blue hexagon designates the Brillouin zone of a single layer of graphene that has gone under either a strain, $\vect \xi_\omega$, or a twist, $\vect \xi_\theta$. In (a) the Brillouin zones are expanded/shrunken with respect to the undeformed Brillouin zone (green dotted hexagon), where as in (b) they are rotated with respect to each other. Each pair gives rise to its moir\'e reciprocal lattice (shown by red hexagonals). Also $K$ points are designated by black dots.}
\label{fig:BZ}
\end{figure}

Consider a bilayer graphene system with a small twist and/or strain applied to the  layers. As emphasized in Ref.~\cite{EmEnSc}, both can be described in terms of a diffeomorphism:  $\vect\xi_\omega \equiv \epsilon\omega (\vec{r}) \vec{r}$ for strain and  $\vect\xi_\theta \equiv \epsilon\theta(\vec{r}) \hat{z} \times \vec{r}$ for a twist (parameterized by $\epsilon \ll 1$). With $\vect\xi_\omega$ and $\vect\xi_\theta$ being orthogonal to each other and thus forming a basis, all  possible flows in two dimensions can be described as a linear combination of these two fields. Therefore, a bilayer system with a general infinitesimal deformation can be achieved by applying the diffeomorphism $\vec{r} \rightarrow \vec{r} + \vect\xi_{\omega,\theta}/2$ to one layer and $\vec{r} \rightarrow \vec{r} - \vect\xi_{\omega,\theta}/2$ to the other, as in Fig.~\ref{fig:BZ}. A finite deformation is reached by successively applying such diffeomorphisms. The time dependent moir\'e fields $\omega(t,\vec{r})$ and $\theta(t,\vec{r})$ can be further looked at as bosonic phonon-like degrees of freedom for the  bilayer system.

Focusing on only one valley, $K$, in the Brillouin zone, an electron that hops from the $K$-point of one layer to that of the other, can do so along three different momentum vectors that will form the moir\'e reciprocal lattice vectors of the bilayer system. This is because the deformation has separated the equivalent $K$-points in three different ways (along $\vec{q}^\theta_{1,2,3}$ for twist and $\vec{q}^\omega_{1,2,3}$ for strain as depicted in Fig. \ref{fig:BZ}) that are related to each other by a $2\pi/3$ rotation. For the general deformation these vectors are given by $\vec{q_1}\equiv 2K_D (\sinh\frac{\omega}{2} \hat{x}-\sin\frac{\theta}{2}\hat{y})$ with $\vec{q}_{2,3}$ derived by successive $2\pi/3$ rotations of $\vec{q}_1$, while $K_D$ is the distance between $K$-points of the undeformed Brillouin zone and its center.

\begin{figure}
\includegraphics[width=\linewidth]{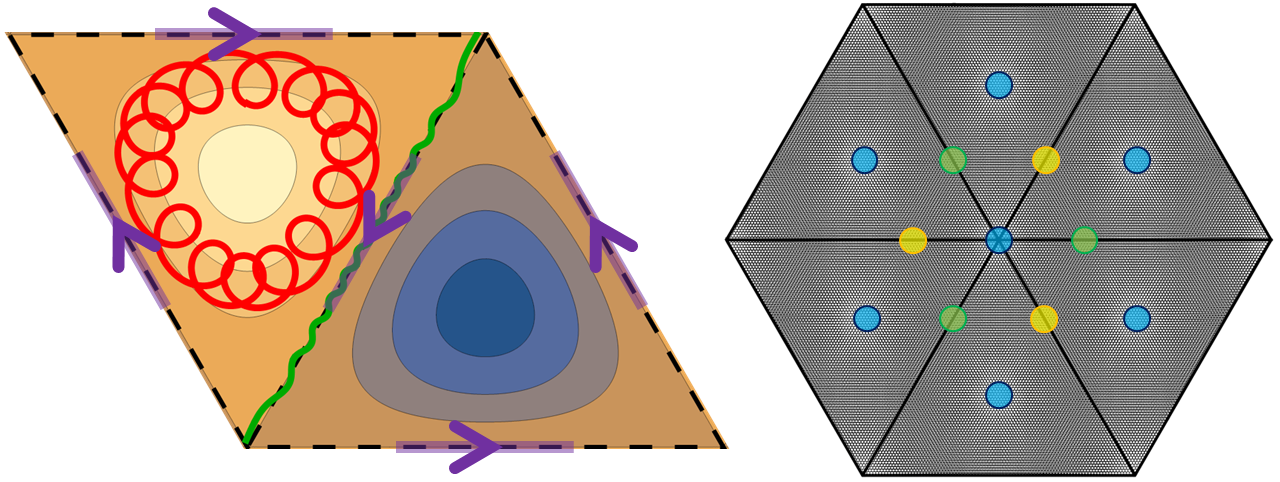}
\centering
\caption{On the left, a generic electron (red trajectory) drifting in the magnetic region while another electron (green trajectory) is moving close to the edge of one magnetic region. A right or left handed fermion belongs only to either of the magnetic regions but since the sides are identified a right handed fermion in one region is the left handed fermion in the other. Having one electron in both regions is similar to having two electrons in one region and forgetting about the other. On the right the twisted bilayer graphene at first magic angle with blue dots denoting AA stacking and yellow/green dots AB/BA stacking. The distance between the neighboring equivalent dots is equal to $L$. }
\label{fig:Drift}
\end{figure}

%The characteristic length of the superstructure, i.e. t
The moir\'e lattice constant, $L$, can be read off of the magnitude of the moir\'e reciprocal lattice vectors, $|\vec{q}|=2K_D(\sinh^2\frac{\omega}{2}+\sin^2\frac{\theta}{2})^{1/2}$, which is equal to the distance between the corresponding $K$-points (denoted by $k$ and $k'$ in Fig. \ref{fig:BZ}): $L=4\pi/3|\vec{q}|$. If the electron hops from one layer to the other without changing its position, it acquires a phase  determined  only by the moir\'e reciprocal lattice vectors. The dynamics of this electron is given by the following Hamiltonian density~\cite{balents},
\begin{align}
    h(\vec{r}) \! = \! \! \left[ \!
    \begin{array}{cc}
     -i\nu \slashed{\partial} & T(\vec{r})  \\
     T^\dagger (\vec{r}) & -i\nu \slashed{\partial}
    \end{array}
    \! \right] \! , \  
    T(\vec{r}) \! \equiv \! \! \left[ \!
    \begin{array}{cc}
     v V(\vec{r}) & u U^*(-\vec{r})  \\
     u U(\vec{r}) & v V(\vec{r})
    \end{array}
    \! \right] \! , 
    \label{HBilayer}
\end{align}
where $-i\nu \slashed{\partial}$ is the Hamiltonian density for a single layer of graphene with $\slashed{\partial}\equiv\sigma^i\partial_i$, and $T(\vec{r})$ is the interlayer tunneling matrix and encodes the periodic profile of the moir\'e pattern, Fig. \ref{fig:Drift}. $V(\vec{r})=\sum_j e^{-i\vec{q}_j \cdot \vec{r}}$ takes the electron from one sublattice (either $A$ or $B$) to the same sublattice on the other layer, while $ U(\vec{r})=\sum_j e^{-i\vec{q}_j \cdot \vec{r}} e^{i (j-1) 2\pi /3 }$ takes it to the opposing sublattice, with $v$ and $u$ being the corresponding tunneling amplitudes and $j=1,2,3$.

Using the algebra of gamma matrices, $\{\gamma^\mu,\gamma^\nu\}=2\eta^{\mu\nu}$, with $\eta_{\mu\nu}$ as the metric, and the unitary transformation,
\begin{align}
    h(\vec{r})\rightarrow \Omega h(\vec{r}) \Omega^\dagger \ \ \ \text{with} \ \ \ \Omega = \frac{1}{\sqrt{2}}\left[
    \begin{array}{cc}
     -\mathds{1} & \mathds{1}  \\
     \sigma^z & \sigma^z
    \end{array}
    \right] \, ,
\end{align}
we can write the Hamiltonian density \eqref{HBilayer} in terms of Dirac fermions coupled to non-fluctuating  gauge fields as follows
\begin{align}
    H = \int \! d^2 x \Big[\, \bar{\psi} i \nu \gamma^a \left( \partial_a + i \mathcal{A}_a \gamma_5 + i\mathcal{S}_a i\gamma_3 \right)\psi \nonumber \\
     -\bar{\psi}\gamma^0 \mathcal{A}_0 \gamma_5 \psi - \bar{\psi}\gamma^0 \mathcal{S}_0 i\gamma_3\psi \Big] \, ,
\end{align}
%\begin{align}
%    H = \int \! d^2 x \, \bar{\psi} i \nu \left( \gamma^a \partial_a + i\gamma^\mu \mathcal{A}_\mu \gamma_5 + i \gamma^\mu  \mathcal{S}_\mu i\gamma_3 \right)\psi
%\end{align}
with $\bar\psi\equiv \psi^\dagger\gamma^0$. Or in the even tidier action formulation,
\begin{equation}
    S = \int \! d^3x \, \bar{\psi} i\slashed{D}\psi  , \quad \! \slashed{D}\equiv \gamma^\mu \left( \partial_\mu + i \mathcal{A}_\mu \gamma_5 + i  \mathcal{S}_\mu i\gamma_3 \right)  , 
    \label{action3D}
\end{equation}
where $a=1,2$, $\mu=0,1,2$ and the field components are,
\begin{align} \label{GFields}
    &\mathcal{A}_0 = -\frac{v}{\nu }\text{Re}[V(\vec{r})], \ \ \mathcal{S}_0 = \frac{v }{\nu }\text{Im} [V(\vec{r})], \\
    & \mathcal{A}_1  = \frac{u }{2\nu }\text{Re} [U(\vec{r})+U(-\vec{r})], \  \mathcal{A}_2  =  \frac{u }{2\nu }\text{Im} [U(\vec{r})+U(-\vec{r})] , \nonumber \\
    &\mathcal{S}_1 =\frac{u }{2\nu }\text{Im}  [U(\vec{r}) - U(-\vec{r})], \ \mathcal{S}_2 =  \frac{u }{2\nu }\text{Re}  [U(-\vec{r})-U(\vec{r})]. \nonumber
\end{align}
The total field strength associated with $\slashed{D}$ is given by $F_{\mu\nu}=\gamma_5 F^\mathcal{A}_{\mu\nu} + i\gamma_3 F^\mathcal{S}_{\mu\nu} + 2i\gamma_5\gamma_3 \mathcal{A}_\mu S_\nu$, with $F_{\mu\nu}^{\mathcal{A},\mathcal{S}}$ being the field strengths generated by $\mathcal{A}_\mu$ and $\mathcal{S}_\mu$ respectively. 

\begin{figure*}
\includegraphics[width=\textwidth]{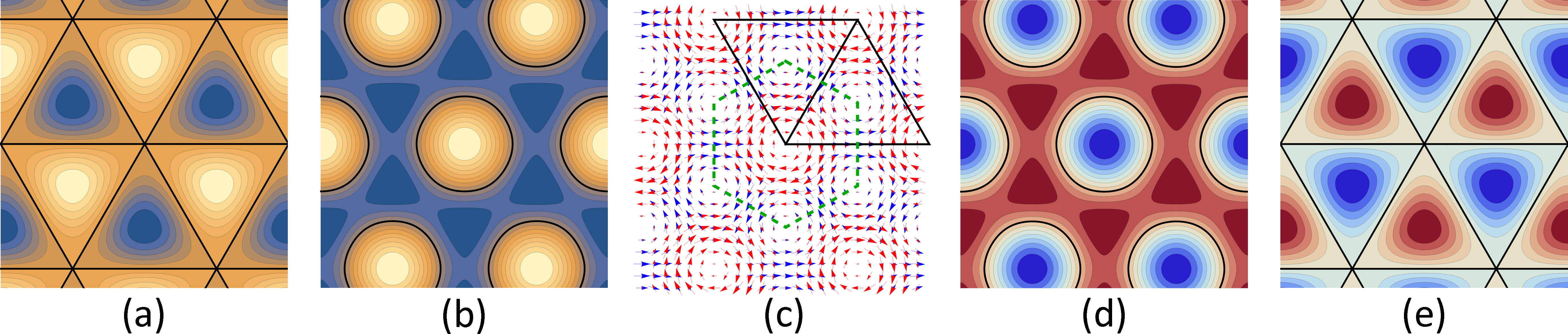}
\centering
\caption{(a) The magnetic field $B$ created by $\mathcal{A}_a$ felt by $\psi_+$ while its negative is exerted upon the $\psi_-$. (b) The field strength associated with the spin field. (c) Vector fields $\mathcal{A}_a$ (blue) and $\mathcal{S}_a$ (red). The black line designates two magnetic regions related by parity. Parallel sides are identified with each other and the whole system can be reconstructed by sewing these together. The $\mathcal{S}_a$ field is zero everywhere on the green dashed hexagon. (d) The chiral scalar potential $\mathcal{A}_0$ as experienced by $\psi_+$ and (e) that of $\mathcal{S}_0$. Except for (c) the fields are zero on the black curves. Note how $\mathcal{S}_0$ coincides with the magnetic field in (a) and also how $\mathcal{A}_0$ coincides with field strength in (b), in particular $\mathcal{S}_0$ vanishes on edge of each magnetic region. Also note that a fermion configuration localized on the edge of each magnetic region will be perpendicular to $\mathcal{S}_a$ (see (c) for example).}
\label{fig:FStrength}
\end{figure*}

Looking at the action \eqref{action3D}, we see that the bilayer problem has transformed into that of Dirac fermions moving in a $(2+1)$ dimensional spacetime and acted upon by two axial-vector fields: A chiral gauge field, $\mathcal{A}_\mu$, and a spin field, $\mathcal{S}_\mu$. Note that $\gamma_3$ measures the spin along the direction normal to the material plane.
The gauge fields, \eqref{GFields}, are periodic with periodicity $\sqrt{3}L$, hence their corresponding field strengths are proportional to $1/L$ and also periodic with the same period, while the distance between each minimum and its neighboring maximum is $L$ (see Fig. \ref{fig:FStrength} for example). In particular the spatial part of the field strength $F^\mathcal{A}_{12}=\partial_2 \mathcal{A}_1 - \partial_1 \mathcal{A}_2$ is given by 
\begin{align} \label{B}
    B(\vec{r}) \equiv F^\mathcal{A}_{12}=\frac{u}{\nu}\sum_j\hat{\vec{q}}^\theta_j \cdot \vect{\nabla} (\vec{q}_j \cdot \vec{r}) \, \sin \! {\left(\vec{q}_j\cdot \vec{r}\right)} \, ,
\end{align}
where $\hat{\vec{q}}^\theta_j \cdot \vect{\nabla}$ is the derivative along the unit vector $\hat{\vec{q}}^\theta_j \equiv \vec{q}^\theta_j/|q^\theta_j|$. That of $\mathcal{S}_a$ equals $\frac{u}{\nu}\sum_j\hat{\vec{q}}^\theta_j \cdot \vect{\nabla} (\vec{q}_j \cdot \vec{r}) \, \cos \! {\left(\vec{q}_j\cdot \vec{r}\right)}$.

Consider the following chiral transformations associated with action \eqref{action3D}
\begin{align}
    \left\{ \!
    \begin{array}{cc}
     \bar\psi \rightarrow \bar\psi e^{i\alpha\gamma_5} \\
     \psi \rightarrow e^{i\alpha\gamma_5}\psi   
    \end{array}
    \! \right\} \! , \ 
    \left\{ \!
    \begin{array}{cc}
     \bar\psi \rightarrow \bar\psi e^{i\alpha i\gamma_3} \\
     \psi \rightarrow e^{i\alpha i\gamma_3}\psi
    \end{array}
    \! \right\} \! , \ 
    \left\{ \!
    \begin{array}{cc}
     \bar\psi \rightarrow \bar\psi e^{i\alpha\Gamma} \\
     \psi \rightarrow e^{i\alpha\Gamma}\psi  
    \end{array}
    \! \right\} \! ,
\end{align}
with $\Gamma \equiv \gamma_0\gamma_3\gamma_5$. Each one of the above chiral transformations can become a symmetry of the action \eqref{action3D} under additional constraints:  $\psi \rightarrow e^{i\alpha\gamma_5}\psi$, $\psi \rightarrow e^{i\alpha i\gamma_3}\psi$, and  $\psi \rightarrow e^{i\alpha\Gamma}\psi$ are promoted to symmetries if $\mathcal{S}_\mu=0$, $\mathcal{A}_\mu=0$, and  $\mathcal{A}_0=\mathcal{S}_0=0$ respectively. Chiral particles can be defined with respect to each of these symmetries by using projection operators, e.g. $\psi_\pm \equiv \frac{1}{2}(1 \pm \gamma_5)\psi$. Since $\gamma_5$ and $\gamma_3$ do not commute we cannot simultaneously create fermions with definite $\gamma_5$ and $\gamma_3$ handedness, in contrast to $\Gamma$ which commutes with both. So we can, for example, have $\psi_{\updownarrows}^{\circlearrowright , \circlearrowleft} \equiv \frac{1}{4}(1\pm i\gamma_3)(1\pm\Gamma)\psi$.

Our interest here is focused, more than anything else, on flat bands, which can be looked at as a class of modes covering the whole Brillouin zone at constant energy, $\bar\psi_k\gamma^0\partial_0\psi_k=\mu\psi^\dagger_k\psi_k$, within which the electrons are therefore localized $\partial E_k / \partial k =0$. If we only consider this class, we eliminate  the time dependence from the action entirely and reduce the $(2+1)$ dimensional theory to its $(2+0)$ dimensional version. Specifically, in the case of $v=0$, which  supports exact flat bands~\cite{Origin}, we have
\begin{equation}
    I = \int \! \mathcal{D} \bar{\psi}\psi \exp\!\left\{\int \! d^2x \,  \bar{\psi} i \gamma^a \left( \partial_a + i \mathcal{A}_a \gamma_5 + i  \mathcal{S}_a i\gamma_3 \right)\psi \right\} \! .
    \label{I2D}
\end{equation}
Using any of the chiral projections, $\psi^{\circlearrowright , \circlearrowleft}_{\pm \ \text{or} \  \updownarrows}$, we can break the above path integral further down, for instance, to
\begin{align}     \nonumber
     I = \int \! \mathcal{D} \bar{\psi}_\pm\psi_\pm   \exp\!\Bigg\{\int \! d^2x \Big[  \bar{\psi}_\pm i \gamma^a \left( \partial_a \pm i \mathcal{A}_a \right)\psi_\pm \\
     \quad \  + \bar\psi_\mp \gamma^a \epsilon_{ab} \mathcal{S}^b    \psi_\pm  \Big] \Bigg\} \,  ,  \label{IL2D}
\end{align}
where the path integral is over the four field variables $\bar{\psi}_\pm$ and $\psi_\pm$, while $\psi_\pm$ fermions are coupled to $\pm \mathcal{A}_a$.

In this form the anomaly residing in the theory given by the path integral \eqref{I2D} takes the familiar shape of the chiral anomaly in two dimensional Euclideanized spacetime. In a path integral such as above the gauge field $\mathcal{A}_a$ has an index associated with it that is directly given by the chiral anomaly~\cite{Fujikawa2004Book,ASIndex,NonText}. But the index must be an integer number which as we will see is only possible for certain values of $\theta$ and $\omega$ that coincide with the magic angle. This consistency condition can therefore tell us whether a flat band exists or not. Before proceeding to a more detailed investigation, it is worth mentioning that this reasoning, following the reduction from Eq. \eqref{action3D} to Eq. \eqref{IL2D}, is  generalizable to other perhaps more complicated systems such as multilayer graphene. In which case we should use higher dimensional gamma matrices to accommodate for the additional layers. Other examples may include Refs. \cite{LiangFu1,LiangFu2,ChouWilson}.

Since the gauge potentials are periodic the path integral can be divided into equivalent patches sewn together by an integration over all field configurations on the boundaries.
\begin{align}
    &I = \int \prod_\tri \mathcal{D}\bar\psi_\dtri\psi_\dtri \, I_\tri \left[\bar{\psi}_\dtri , \psi_\dtri \right] \, , \\
    & \text{with} \quad I_\tri \left[\bar{\psi}_\dtri , \psi_\dtri \right]= \int_{\bar\psi_\dtri,\psi_\dtri} \mathcal{D}\bar\psi_\tri\psi_\tri e^{iS_\tri} \, ,
\end{align}
where $\dtri$ designates the boundary of the patches and the configuration residing on it, while $\tri$ designates the patch itself. $I_\tri$ is the path integral over all configurations on one patch, $[\bar\psi_\tri,\psi_\tri]$, that go to $[\bar\psi_\dtri,\psi_\dtri]$ on the boundary. Also $S_\tri$ is the same action before but with an integral only over $\tri$. The patches are chosen so that the action $S_\tri$ is the same in all $I_\tri$. While satisfying this property, we choose $\tri$ so that $\mathcal{S}_a$ will either vanish on or be perpendicular to $\dtri$. This way the edge configurations along $\dtri$ will have an additional chiral symmetry with respect to $e^{i\alpha\gamma_5}$. 

If the fermions remain confined within one patch, which should be the case when they are localized, we can exclude from the path-integration those configurations that connect different patches together. The probability density and current are here given by $\psi^\dagger\psi$ and $j^a \equiv \bar\psi\gamma^\mu\psi$ respectively. We expect the excluded configurations to be those with a nonzero flow of probability current, $j^a$, out of $\tri$: $\int_\tri \partial_0 \psi^\dagger \psi=\int_\tri \partial_a j^a = \int_\dtri \hat{n}_\dtri^a \cdot j_a  \neq 0$. In that case the problem is reduced, from the initial path integral $I$, to segregated path integrals of $I_\tri=\int \mathcal{D}\bar\psi_\tri\psi_\tri \exp\{iS_\tri\}$. If, moreover, there is a flat-band then the transition amplitude and thus the path integral, from any state to any other state within the band would be time independent
\begin{equation}
    \bra{\zeta}e^{iHt}\ket{\chi} \equiv \int^\zeta_\chi \mathcal{D}\bar{\psi}\psi e^{iS} = \braket{\zeta}{\chi} \, .
\end{equation}
This allows us to reduce the theory to $(2+0)$ dimensions as in Eq. \eqref{action3D} to Eq. \eqref{IL2D}:
\begin{align}     \nonumber
     I_\tri = \int_\tri \! \mathcal{D} \bar{\psi}_\pm\psi_\pm   \exp\!\Bigg\{\int_\tri \! d^2x \Big[  \bar{\psi}_\pm i \gamma^a \left( \partial_a \pm i \mathcal{A}_a \right)\psi_\pm \\
     \quad \  + \bar\psi_\mp \gamma^a \epsilon_{ab} \mathcal{S}^b    \psi_\pm  \Big] \Bigg\} \,  ,  \label{IR2D}
\end{align}
where we have removed the $\tri$ sign from the fermionic field variables (and brought it under the path integral sign instead) to avoid clutter. The most important role of the spin field $\mathcal{S}_a$ is that it takes $\psi_\pm$ to $\psi_\mp$ which is required for an intact patching of neighboring $\tri$s~\cite{ZeroFlux}. Otherwise wave-functions leak and the flat-band gains a curvature. Effectively one can translate this into periodically changing the sign of the gauge field $\mathcal{A}_a$ so that its field strength will not average to zero. For the moment we assume this role played and refer the reader to Supplemental Material for the detailed discussion.

Focusing only on $I_\tri$, we see however, that not all gauge field configurations fit within the boundaries $\dtri$; only those with a complete integer index, $n_\circlearrowright-n_\circlearrowleft\in\mathbb{Z}$. One way to observe this is first to notice that a continuous chiral rotation of $\psi_+ \rightarrow e^{i2\pi\Gamma}\psi_+=\psi_+$ takes the spinor field to itself while leaving the action unchanged. However, the theory \eqref{IR2D} is subject to chiral anomaly, namely, the Jacobian of our chiral transformation, $J_5$, is non-trivial,
\begin{align}     \label{Rot2Pi}
    I_\tri = &\int_\tri \! \left[ \mathcal{D} \bar{\psi}_+\psi_+ \right] \mathcal{D} \bar{\psi}_-\psi_- e^{iS_\tri}  \longrightarrow \\
    &\int_\tri \! \left[\mathcal{D} \bar{\psi}_+\psi_+  J_5 \right] \mathcal{D} \bar{\psi}_-\psi_- e^{iS_\tri}= I_\tri e^{i 2\pi(n_\circlearrowright - n_\circlearrowleft)} \, . \nonumber
\end{align}
The last equality above comes from knowing that the Jacobian of chiral transformation is connected to the Atiyah-Singer index~\cite{Fujikawa,Fujikawa2004Book,ASIndex}. Chiral transformation discriminates between right and left handed modes, $n_\circlearrowright - n_\circlearrowleft$, hence the path integral (which yields the determinant of the Dirac operator) obtains a phase associated with this difference. This phase encodes the winding number of the gauge field associated with the Dirac operator---an integer number which in our two dimensional system is written as,
\begin{equation}
    \frac{1}{2\pi}\int_\tri d^2x \, \epsilon^{ab}\partial_a \mathcal{A}_{b} = n_\circlearrowright - n_\circlearrowleft \, .
\end{equation}

But since the field variables do not change by a complete rotation, in Eq. \eqref{Rot2Pi}, the path integral must also remain the same: $I_\triangledown = I_\triangledown e^{i2\pi(n_\circlearrowright - n_\circlearrowleft)} $. Thus, if $e^{i2\pi(n_\circlearrowright - n_\circlearrowleft)}\neq 1$, then the only way that the initial and the transformed path integrals can be equal is for them to vanish, $I_\tri=0$. This zero valued partition function means that the state is unrealizable~\footnote{In contrast, the flat-band can be realized if $n_{\circlearrowright,\circlearrowleft}\in\mathbb{Z}$.}.

What we have discussed so far applies generally to all deformations of any bilayer system that shares the symmetries of graphene. Let us now focus on uniform twist for which $\vec q_1=\vec q_1^\theta=q^\theta (-1,0)$ and $\vec q_{2,3}=\vec q_{2,3}^\theta=q^\theta (\pm\sqrt{3}/2,1/2)$ with $q^\theta = 2K_D \sin (\theta/2) = 4\pi/3L$. The gauge fields $\mathcal{A}_a$ and $\mathcal{S}_a$ generated by uniform twist are divergenceless with their corresponding field strengths proportional to $u/\nu L$.
%The gauge potentials and field strengths are shown
These are depicted in Fig. \ref{fig:FStrength}.
We need to calculate the minimum value of $L$ corresponding to $n_\circlearrowright = 1$.
%Using Fujikawa's method of calculating anomalies~\cite{Fujikawa2004Book} we find,
We find,
\begin{align}
    n_\circlearrowright \! = \! \frac{1}{4\pi}\int_\triangledown \! d^2x  \, \epsilon^{ab}\partial_a \mathcal{A}_b  = \frac{1}{4\pi}\int_\triangledown \! d^2x \, B  = \frac{3\sqrt{3} u }{4\pi\nu} L  \, ,
    \label{IndexR}
\end{align}
which is equal to $1$ for $L=L_0\equiv 4\pi\nu/3\sqrt{3}u$. Considering $L=a /2\sin(\theta/2)$, where $a=2.46 \text{\AA}$ is the graphene lattice constant, the first magic angle is given by $\theta \approx 3\sqrt{3} au/4\pi \nu \approx 1.1^\circ$ with $u=0.11 \text{eV}$ and $\nu K_D = 9.9 \text{eV}$.

To develop semi-classical intuition, let us first  assume that the spin current term $\mathcal{S}_a \bar\psi\gamma^a\gamma_3\psi$ is disregardable. Then we are left with only a magnetic field with strength proportional to $u/\nu L$ acting with opposite signs on $\psi_\pm$ fermions that are completely decoupled from each other. If the magnetic field was constant across the material then the electrons would have been subject to Landau localization rotating around a fixed center, in a semi-classical picture, and forming Landau levels. Since the moir\'e effective magnetic field is inhomogeneous, the semi-classical picture changes to that of drifting fermions---rotating around a cycling center. See Fig. \ref{fig:Drift}. For localization to be possible, the drifting fermion should be able to fit into one magnetic region. The smallest rotating fermion, according to the uncertainty principle, has a size of $\ell_B \propto 1/\sqrt{\bar B}$ and, since here the average magnetic field $\bar{B}$ is proportional to $1/L$, it expands with $\sqrt{L}$. But the size of the magnetic regions, or $\tri$, is proportional to $L$ which grows faster than $\ell_B\propto \sqrt{L}$. Thus $\tri$ gets bigger faster than the smallest possible electron and eventually can catch one, at which moment one electron has just been trapped inside the magnetic region and can complete a cycle there without getting out of it.

At that moment we expect to have an edge mode on the boundary of the magnetic region. Appropriately, for this mode the effect of spin current term on the magic angle is indeed disregardable since $\mathcal{S}_a$ and $\bar\psi\gamma^a\gamma_3\psi$ are perpendicular to each other on $\dtri$. But if there is an edge mode it means that the number of right or left handed fermions that reside in the magnetic region must at least be one, $n_{\circlearrowright,\circlearrowleft}=1$. This again leads us to Eq. \eqref{IndexR} and the magic angle. It also can be seen in terms of unit of flux and a restricted type of Landau quantization. The size of the unit flux is given by $\ell_B=\sqrt{L \nu/u}$. Therefore, $n=L^2/2\pi \ell_B^2 = uL/2\pi\nu$ will be the degeneracy of the \textit{moir\'e Landau level}, at least for large $n\in \mathbb{Z}$. Thus we expect a series of magic angles connected to each other by steps of $\delta L \approx 2\pi \nu/u$~\footnote{In terms of the parameter $\alpha$ introduced in Ref.~\cite{Origin} these are steps of $\delta \alpha \approx 3/2$ while the first magic angle corresponds to $\alpha \approx 1/\sqrt{3}$.}.

The negligibility of $\mathcal{S}_a$ for the first magic angle demonstrates that the non-Abelian theory can be \textit{Abelianized} for this case. For a similar process see Ref.~\cite{NonAbelianBos} where a spin field, $S_\mu$, is decoupled from the fermions. In turn, it means that the physics of the first magic angle maps to that of the lowest Landau level.
%including the interactions and variuos fillings
For higher magic angles, as the zero-mode wave-functions become more complicated, the effects of $\mathcal{S}_a$ also become considerable. Since the field strength is given by $F_{\mu\nu}=\gamma_5 F^\mathcal{A}_{\mu\nu} + i\gamma_3 F^\mathcal{S}_{\mu\nu} + 2i\gamma_5\gamma_3 \mathcal{A}_\mu S_\nu$, the phase associated to non-Abelian part grows with $L^2$ while that of the Abelian part grows with $L$. Hence, Abelianization is plausible for $L \lesssim 4\pi\nu/3u$. For a longer discussion including the definition of Abelianization and few implications such as the first order correction due to $\mathcal{S}_a$, see the Supplemental Material.
%The first order correction coming from $\mathcal{S}_a$ takes us from $\alpha \approx 0.577$ to $\alpha \approx 0.589$.
A more detailed consideration is given in an upcoming paper~\cite{ZeroFlux}. Also look at Ref.~\cite{ParhizkarLocalizing} for a general discussion.

Reformulating the theory in terms of Dirac fermions \eqref{action3D} has extra merits. For example, the application of an external electromagnetic field yields the same theory  \eqref{action3D} but now with the Dirac operator carrying an additional external gauge field, $A_\mu$,
$$\slashed{D}\equiv \gamma^\mu \left( \partial_\mu + i A_\mu + i \mathcal{A}_\mu \gamma_5 + i  \mathcal{S}_\mu i\gamma_3 \right) \,.$$
 Upon projecting the Dirac fermions into $\psi_\pm \equiv \frac{1}{2}(1\pm\gamma_5)\psi$ as before, we see that the chiral fermions  are now coupled to the shifted gauge fields $A_a \pm \mathcal{A}_a$. For example, if $A_a$ is due to a constant magnetic field $H$, Eq. \eqref{IndexR} reads
\begin{align} 
    n_\circlearrowright  = \frac{1}{4\pi}\int_\triangledown \! d^2x \, \left( H \pm B \right)  = \frac{H}{4\pi}\frac{3\sqrt{3}}{4 } L^2 \pm \frac{3\sqrt{3} u }{4\pi\nu} L \, .
    \label{IndexRH}
\end{align}
For $u^2/\nu^2 > 4\pi H / 3\sqrt{3}$, the requirement $n_\circlearrowright=\pm 1$ has more than one solution in contrast to Eq. \eqref{IndexR}. Thus, an external magnetic field splits each magic angle into
\begin{equation} \label{LH}
    L = \pm \frac{u}{\nu}\frac{2}{H} \pm \sqrt{\left(\frac{u}{\nu}\frac{2}{H}\right)^2 \pm \frac{16\pi}{3\sqrt{3}H}} \, ,
\end{equation}
with the magic \textit{angles} given by $\theta = 2\arcsin (a/2L)$, while each combination of pluses and minuses above yields a solution to Eq. \eqref{IndexRH}.
For a small external magnetic field $H \ll (3\sqrt{3}/4\pi) u^2/\nu^2 \approx 140 \text{mT}$ %(or in terms of its corresponding magnetic length $\ell_H \gg 15 a$)
the positive solutions can be written as,
\begin{equation}
    L_1^\pm = L_0 \pm \frac{4\pi^2}{27}\frac{\nu^3}{u^3}H \, , \ \ L_2^\pm= \frac{4}{H}\frac{u}{\nu}  \pm L_0 -\frac{4\pi^2}{27}\frac{\nu^3}{u^3}H  \, .
\end{equation}
with $L_0 \equiv 4\pi\nu/3\sqrt{3}u$, being the magic angle in the absence of the external magnetic field. In the limit $H\rightarrow 0$ we regain the previous magic angle from $L_1^\pm$, in addition to having $L_2^\pm \rightarrow \infty$ that happens when the bilayer is untwisted, $\theta \rightarrow 0$, and the moir\'e reciprocal lattice, which would have a vanishing size, is in fact flat. The dispersion at this instance is that of two quadratic bands touching at the $K$-point; or in other words a vanishing Dirac velocity.
%In the large $H$ limit, the external magnetic field will be dominant and confined flux will be realized for smaller values of $L$.

Let us also consider applying a uniform strain on top of the already existing uniform twist. The moir\'e pattern will rotate by $\arccos(q^\theta/|\vec{q}|)$ and its length, $L$, will shrink accordingly. In this case, using Eq. \eqref{B}, the magnetic field generated by $\mathcal{A}_a$ and its corresponding index are given by $B=q^\theta \frac{u}{\nu}\sum_j \sin(\vec q_j\cdot \vec r)$ and $n_\circlearrowright=(q^\theta/|\vec{q}|)L/L_0$ respectively. Comparing this with Eq. \eqref{IndexR} we see that the flat-band now happens at,
\begin{equation} \label{TwistStrain}
    L=L_0 \sqrt{1 + \frac{\sinh^2(\omega/2)}{\sin^2(\theta/2)}}\, .
\end{equation}

Now consider the case of finite $AA$ hoppings by gradually increasing $v$ from zero. This is equivalent to reintroducing $\mathcal{S}_0$ and $\mathcal{A}_0$ to the Lagrangian. The former vanishes on the edges of magnetic regions (see Fig.~\ref{fig:FStrength}) and therefore will have no effect on the edge mode. On the other hand, the $\mathcal{A}_0$ term acts as an electric potential for either $\psi_\pm$. An edge mode trajectory is perpendicular to constant $\mathcal{A}_0$ curves, so even though $\mathcal{A}_0$ might redistribute the density of the mode along the edge, it will do little to deviate it out of the edge. Therefore, the first magic angle is robust against a non-vanishing $v$. Like a mass term, $\gamma^0\mathcal{A}_0$ does not anti-commute with the generator of the chiral transformation, $\Gamma$. Since we have $\{\slashed{D},\Gamma\} \propto v$ the $v \rightarrow 0$ ensures that the action is invariant under $\psi \rightarrow e^{i\alpha\Gamma}\psi$ and that for each positive eigenvalue of $\slashed{D}$ there is a negative one with the same magnitude. Deviating form the $v=0$ limit breaks particle-hole symmetry, shifts the eigenvalues of $\slashed{D}$, disconnects the number of left- and right-handed zero modes $n_{\circlearrowright,\circlearrowleft}$ from the Jacobian of the transformation, and gives the formerly flat-band a curvature. In this situation we can define the magic angle, where the flatness of the band is approximate, through the flux felt by $\psi_\pm$, e.g. $\int_\tri  F_{12}/4\pi = \int_\tri \mathcal{A}_a \bar\psi_\dtri\gamma^a\psi_\dtri = 1$, but this implies a perfectly flat-band upon the existence of a well-defined $n_{\circlearrowright,\circlearrowleft}$ which can be found only at $v=0$.

As we have seen in this Letter, although anomalies are not present in all dimensions, it is still possible to conjure them in specific situations. In particular, a flat band can be described through an anomaly in the timeless version of its hosting theory. We saw that the dimensionally reduced theory and thus the flat band are not always realizable. In the case of  bilayer graphene, the obstruction comes from the chiral anomaly and the need to satisfy an index condition, which in turn confirms the topological nature of the flat band. The Dirac field theory form, Eq.~\eqref{action3D}, of the bilayer moir\'e lattice problem allows many  generalizations including the finite temperature case, the presence of complex inhomogeneous external fields and general deformations, and interaction effects in the spirit of Refs.~\cite{ChiralAnomalyInt,NonAbelianBos,ParhizkarPath} where the interplay of anomaly with interactions are discussed. Of particular interest are quantum Hall phenomena  and unconventional superconducting pairing associated with the moir\'e gauge fields~\cite{ParhizkarLocalizing}. 

This work was supported by the National Science Foundation under Grant No. DMR-2037158, the U.S. Army Research Office under Contract No. W911NF1310172, and the Simons Foundation.

\bibliography{main}

\begin{thebibliography}{26}
\bibitem [{Non()}]{NonText}%
  \BibitemOpen
  \href@noop {} {\bibinfo {title} {A complete consideration must take the
  non-abelian chiral anomaly into account.}}\BibitemShut {Stop}%
\end{thebibliography}

\end{document}

\title{A Generic Topological Criterion for Flat Bands in Two Dimensions
\\
Supplemental Material}
 
\author{Alireza Parhizkar and Victor Galitski}
    \affiliation{Joint Quantum Institute, Department of Physics, University of Maryland, College Park 20742}

\maketitle

The negligibility of $\mathcal{S}_a$ for the first magic angle demonstrates that the non-Abelian theory can be \textit{Abelianized} for this case. In turn, it means that the physics of the first magic angle maps to that of the lowest Landau level. For higher magic angles as the zero-mode wave-functions become more complicated the effects of $\mathcal{S}_a$ also become considerable. Here we provide an explanation of why the spin field $\mathcal{S}_a$ can be disregarded for the first magic angle along with few other implications.

Since we are only considering the twisted bilayer graphene in this discussion, the theory has only one free parameter designated by $\alpha$ in Refs~\cite{MacDonald,Origin}. We are going to rescale the theory as in Ref~\cite{Origin} so that the dependency on $\alpha$ is more explicit. This is done by setting $|\vect{q}|=1$ or $L=4\pi/3$ and $u/\nu = \alpha$. This way the gauge fields and the corresponding field strengths both depend linearly on $\alpha$.
Since the total field strength is given by $F_{\mu\nu}=\gamma_5 F^\mathcal{A}_{\mu\nu} + i\gamma_3 F^\mathcal{S}_{\mu\nu} + i[\mathcal{A}_\mu\gamma_5,S_\nu \gamma_3]  $, the phase associated to the non-Abelian part grows with $\alpha^2$ while that of the Abelian part grows with $\alpha$. Therefore, Abelianization should be possible for $\alpha \lesssim 1$. 

In more accurate words, the Abelianization of a non-Abelian theory is the process that aligns the spinor field across the space. So if we start with the path integral below,
\begin{equation}
    I = \int  \mathcal{D} \bar{\psi}\psi \exp\left\{\int  d^2x   \bar{\psi} i \gamma^a \left( \partial_a + i \mathcal{A}_a \gamma_5 + i  \mathcal{S}_a i\gamma_3 \right)\psi \right\}  \, ,
    \label{Init}
\end{equation}
we observe that since $\mathcal{A}_a$ and $\mathcal{S}_a$ are in general two different functions of space, the spinor field, with one component along $\gamma_5$ and another along $i\gamma_3$, rotates by moving across the space. The Abelianized version of the theory above, if it exists, should have the following form,
\begin{equation}
    I = \int  \mathcal{D} \bar{\tilde\psi}\tilde\psi J \exp\left\{\int  d^2x   \bar{\tilde\psi} i \gamma^a \left( \partial_a + i \mathcal{V}_a w_i \gamma^i \right)\tilde\psi \right\}  \, ,
    \label{IAb}
\end{equation}
with $w_{i=\{0,3,5\}}$ a constant direction in the spinor space, $\bar{\tilde\psi}\bar{U} \equiv \bar\psi$ and $U\bar\psi \equiv \psi$ being the rotated fermionic degrees of freedom, and $J$ the Jacobian of this rotation. Keep in mind that $\psi$ and $\bar\psi$ are independent variables in the path integral formulation. The Jacobian has an important role to play, however, the main purpose of this supplement is to investigate whether the theory is Abelianizable or not and here we do not concern ourselves with the Jacobian, regarding to which the reader is encouraged to read Ref.~\cite{NonAbelianBos} where we completely decouple a spin field from fermionic degrees of freedom.
\begin{figure}
\includegraphics[width=\textwidth]{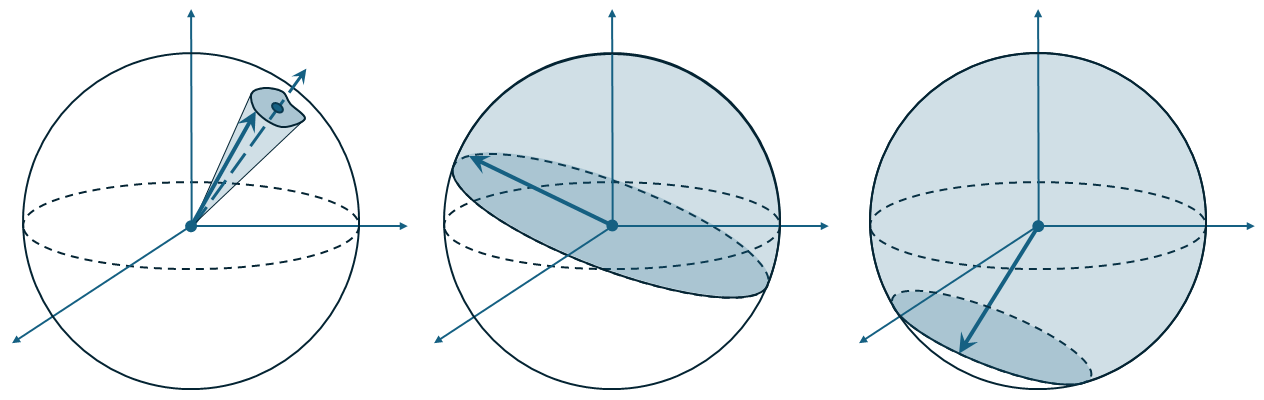}
\centering
\caption{If the span of the trajectory of the spinor is small enough, as it is in the left figure, the theory is Abelianizable.}
\label{fig:Abelianization}
\end{figure}
The above (i.e., translating Eq.\eqref{Init} into Eq.~\eqref{IAb}) is the general definition of the Abelianization, which is however not generally a trivial task. Further we can specifically demand what $U$ should look like and what properties it should preserve. In terms of $U$ and $\bar U$ we want,
\begin{equation}
    \bar{U} \slashed{D} U  \equiv \bar{U} \left( \slashed{\partial} + i \slashed{\mathcal{A}} \gamma_5 + i  \slashed{\mathcal{S}} i\gamma_3 \right) U =  \slashed{\partial} + i \slashed{\mathcal{V}} w_i \gamma^i \equiv \slashed{D}_A \, .
\end{equation}
upto a coordinate transformation, where Dirac slash notation, $\slashed{C}\equiv \gamma^\mu C_\mu$, has been employed. For example, if the transformation also includes a rotation along $\hat{x}$ direction, which is the Lorentz transformation that combines $y$ and $z$ dimensions, it therefore carries the commutator of the corresponding gamma matrices $[\gamma_2,\gamma_3]$. We can then retract the Lorentz transformation in the vector basis by a coordinate change, $ \frac{dx'^\mu}{dx^\nu} = \Lambda^\mu_\nu(x)$ ,
where $\Lambda^\mu_\nu(x)$ is the usual local Lorentz transformation. This will of course rotate all other vector fields as well including the vector potentials, $\mathcal{A}_\mu \rightarrow \Lambda^\nu_\mu \mathcal{A}_\nu$ .
In the cases where there is a $\pi$ rotation and $\hat{z} \rightarrow -\hat{z}$, the resulting two-dimensional system has gone under a parity and the magnetic field resulting from $\mathcal{A}_a$ has reversed.
On the other hand, $[\gamma_2,\gamma_3]$ does not commute with $\gamma_3$, thus the rotation has an additional cost on the $\mathcal{S}_a$ field. This means that we can decouple the $\mathcal{S}_a$ field gradually while we decrease the flux of the magnetic field generated by $\mathcal{A}_a$. Since flux is proportional to $L$, this in turn means that a larger $\alpha$ is required for the magic angle. We postpone a detailed discussion to an upcoming work and resort to approximations of the twisted bilayer graphene case. 

The moir\'e vector potentials of twisted bilayer graphene are here presented in a way that their divergence vanishes. We also know from Helmholtz theorem that any two-dimensional vector field can be decomposed into its divergence-free and curl-free parts. So the moir\'e vector potentials can be written only in terms of divergence-free fields: $\mathcal{A}_a = \alpha\epsilon_{ab}\partial^b \phi_{\mathcal{A}}$ and $\mathcal{A}_a = \alpha \epsilon_{ab}\partial^b \phi_{\mathcal{S}}$, with $\epsilon_{ab}$ being the totally antisymmetric tensor in two-dimensions. The potentials $\phi_{\mathcal{A},\mathcal{S}}$ are explicitly given by,
\begin{equation}
    \phi_{\mathcal{A}}(\vec{r}) =  -\sum_j \sin (\vec{q}_j\cdot \vec{r}) \, , \quad\quad \phi_{\mathcal{S}}(\vec{r}
) =  -\sum_j \cos (\vec{q}_j\cdot \vec{r}) \, .
\end{equation}
with $\vec{q}_{j=1,2,3}$ being momentum unit vectors along $-\hat{y}$ and its successive $2\pi/3$ rotations. Thus Eq.~\eqref{Init} for the twisted bilayer can be written as,
\begin{equation}
    I = \int  \mathcal{D} \bar{\psi}\psi \exp\left\{\int  d^2x   \bar{\psi} i \gamma^a \left( \partial_a + i \alpha\partial_a \phi_{\mathcal{A}} \Gamma \gamma_5 + i  \alpha\partial_a \phi_{\mathcal{S}} \Gamma i\gamma_3 \right)\psi \right\}  \, ,
\end{equation}
where we remember that $\Gamma \equiv \gamma_0\gamma_3\gamma_5$ commutes with $\gamma_0$, $\gamma_3$, $\gamma_5$, and anti-commutes with $\gamma_1$, $\gamma_2$. Also note that $\gamma^b\Gamma=\gamma^a \epsilon^b_{\ a}$. Let us now apply the following rotation,
\begin{equation}
    \bar\psi \rightarrow \bar\psi e^{-\xi\alpha \phi_\mathcal{S} \Gamma \gamma_3} \, , \quad\quad \psi \rightarrow e^{\xi\alpha\phi_\mathcal{S} \Gamma\gamma_3} \psi \, .
    \label{URot}
\end{equation}
After this rotation the path integral above turns into,
\begin{equation}
    I = \int  \mathcal{D} \bar{\psi}\psi J_\xi \exp\left\{\int  d^2x   \bar{\psi} i \gamma^a \left( \partial_a + i \alpha\partial_a \phi_{\mathcal{A}} \cos (2\xi\alpha\phi_\mathcal{S}) \Gamma \gamma_5 + i[  \alpha(1-\xi)\partial_a \phi_{\mathcal{S}} + i\alpha\partial_a \phi_\mathcal{A} \sin(2\xi\alpha\phi_\mathcal{S})\gamma_5] \Gamma i\gamma_3 \right)\psi \right\}  \, .
\end{equation}
The term inside the brackets could vanish as $\xi \rightarrow 1$, if it would not have been for the presence of $\gamma_5$. When the rotation is large enough, the $\gamma_5$ term resurrects the non-Abelian part that should have been decoupled; the spin does not straightly align along $\gamma_5$ direction but instead precess around it. (See Fig.~\ref{fig:Abelianization}.) However for small enough values of $\alpha$ the $\gamma_5$ term is quadratic in $\alpha$ and therefore disregardable. As $\alpha$ gets larger the non-Abelian part starts to rotate away in the spinor space and we are going to need a rotation in a corrected direction to remove it. For larger and larger $\alpha$s more and more successive rotations are required to be able to decouple the non-Abelian part completely. The fact that for small $\alpha$ the theory is Abelianizable is important since it justifies the disregardability of the spin field and also maps the corresponding flat band to that of the lowest Landau level.

\begin{figure}
\includegraphics[width=\textwidth]{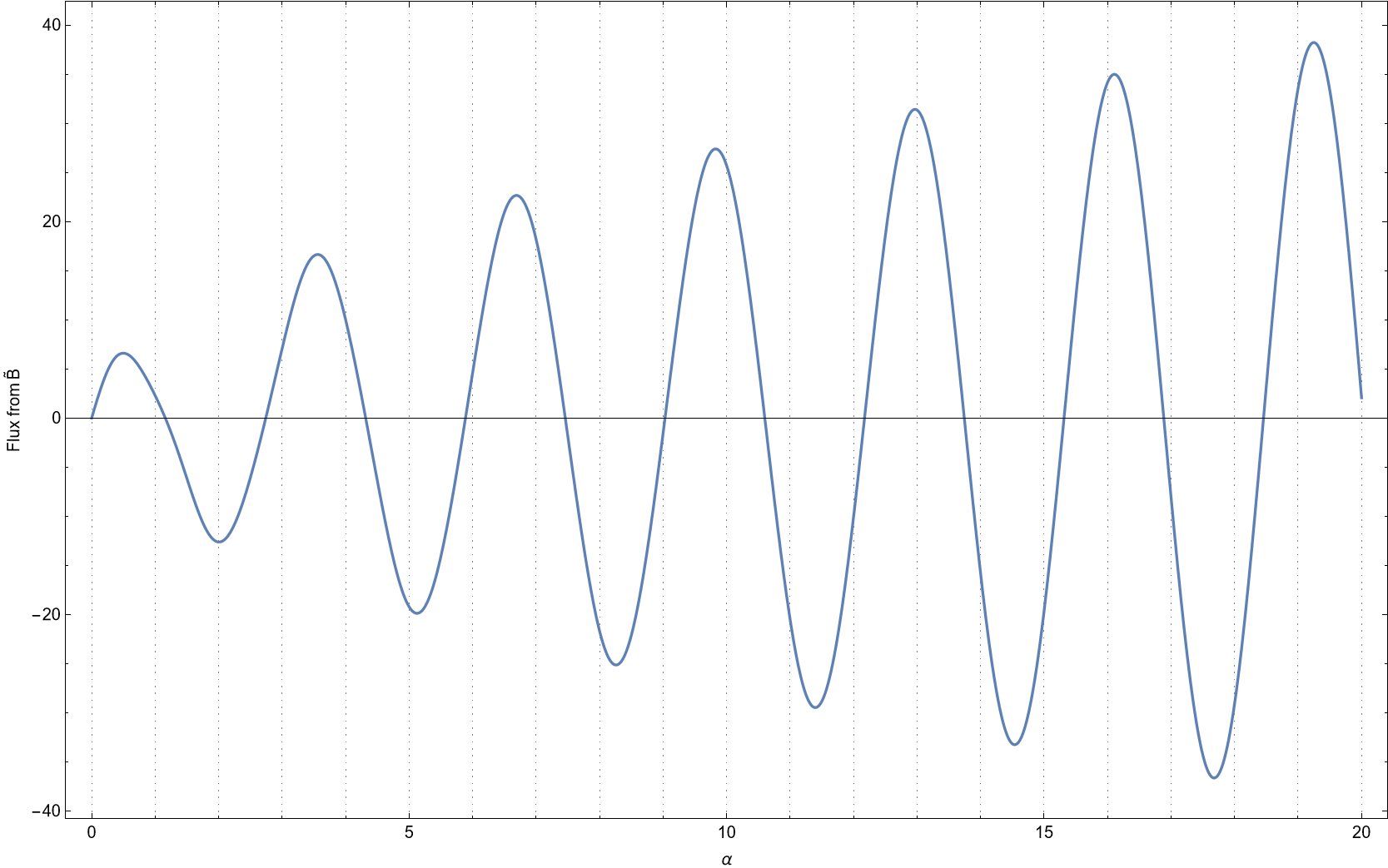}
\centering
\caption{We can use the approximation to estimate the occurrence rate of the magic angles.}
\label{fig:Zeros}
\end{figure}

A true Abelianization is going to correspond to a transformation much more complicated than Eq.~\eqref{URot}. However as an approximation let us apply the above rotation completely by $\xi \rightarrow 1$. We then have,
\begin{equation}
    I = \int  \mathcal{D} \bar{\psi}\psi J_\alpha \exp\left\{\int  d^2x   \bar{\psi} i \gamma^a \left( \partial_a + i \alpha\partial_a \phi_{\mathcal{A}} \cos (2\alpha\phi_\mathcal{S}) \Gamma \gamma_5 + \alpha\partial_a \phi_\mathcal{A} \sin(2\alpha\phi_\mathcal{S}) \Gamma i\gamma_3 \gamma_5 \right)\psi \right\}  \, .
\end{equation}
Now our previous magnetic field $B \equiv \partial_x \mathcal{A}_y - \partial_y \mathcal{A}_x$ is substituted by 
\begin{equation}
    \tilde{B} \equiv \partial_x [\alpha\partial_y \phi_\mathcal{A}\cos(2\alpha\phi_\mathcal{S})] - \partial_y [\alpha\partial_x \phi_\mathcal{A}\cos(2\alpha\phi_\mathcal{S})] = B \left[\cos(2\alpha\phi_{\mathcal{S}}) -
    2\alpha\sin(2\alpha\phi_{\mathcal{S}})\left(\frac{3}{2} + \phi_{\mathcal{S}}\right) \right]\, .
\end{equation}
We have discussed that the field strength flux is giving rise to magic angles. The flux of the magnetic field above is periodically changing with $\alpha$. For large $\alpha$ we can obtain the periodicity by looking at zeros of $\tilde{B}$. These happen roughly when $\cot(2\alpha)=2\alpha \mathcal{C}$ with $\mathcal{C}$ being some number. Since $\cot(2\alpha)$ is divergent for $\alpha = n \pi/2 \approx  3n/2$ the value of $\mathcal{C}$ does not heavily alter the value of $\alpha$ for which $\cot(2\alpha)=2\alpha \mathcal{C}$ is satisfied. Therefore, the periodicity of the flux and thus the occurrence rate of magic angles for large $\alpha$ is roughly given by $\delta\alpha \approx 3/2$. A numerical evaluation of the flux confirms this analysis in Fig.~\ref{fig:Zeros}. It should be emphasized that this is just a heuristic estimation based on the approximation above.

A correction to the value of the first magic angle, without disregarding the $\mathcal{S}_a$ field, can also be estimated. We can ask how far the original flux can be reduced before the non-Abelian precession starts increasing the flux. The answer comes from comparing the original flux generated by $B$ with the flux of $\tilde{B}$. This takes us to the better approximation of $\alpha\approx 0.588$. A more rigorous calculation is left for an upcoming work.

Let us, however, consider another approach to Abelianization by choosing a slightly different $U$ and $\bar{U}$. We know that the spin field $\mathcal{S}_a$ does two things~\cite{ZeroFlux}, first, it effectively gives a preferred frame in which $B$ does not average to zero; second, since it is non-vanishing across the a $\tri$ region it slightly changes the value of the magic angle.
\begin{figure}
\includegraphics[width=\textwidth]{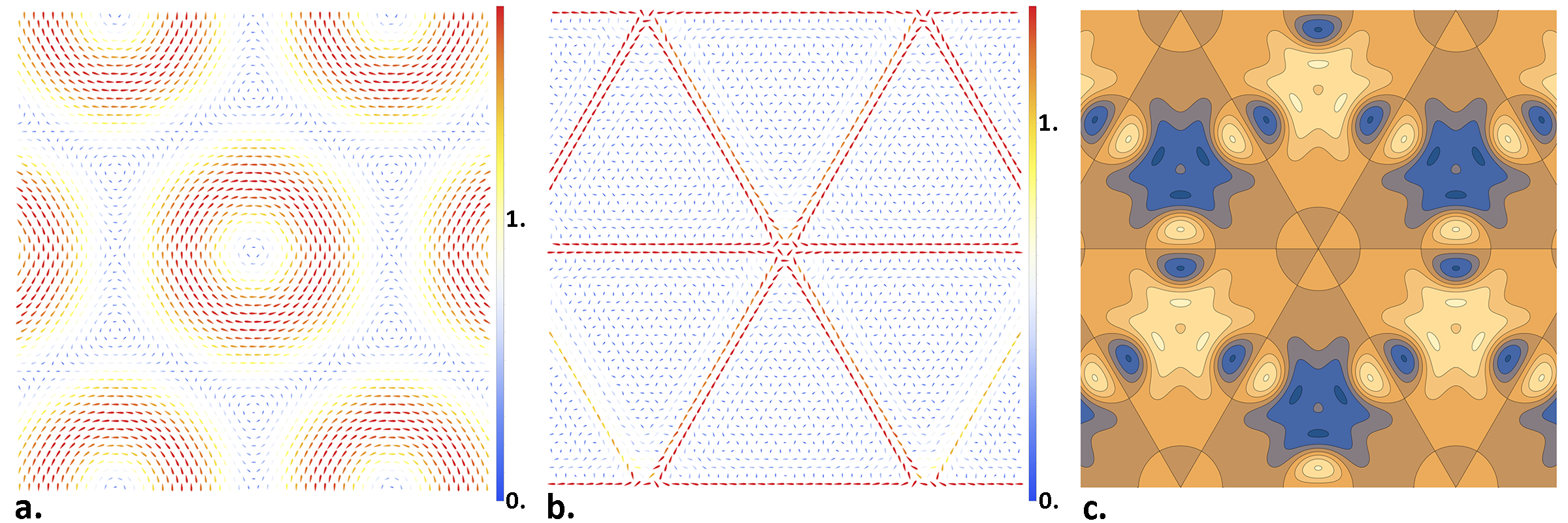}
\centering
\caption{By successive rotations we can turn the spin field on the left (a) to the one in the middle (b) where the spin field is zero almost everywhere except at the boundaries of $\tri$ region. The right figure depicts the magnetic field $\bar{B}$ at $\alpha\approx 0.589$ after only one such rotation (compare to Fig. 2-a in the main text).  }
\label{fig:Fields}
\end{figure}
\begin{equation}
    \bar\psi \rightarrow \bar\psi e^{\alpha \phi_\mathcal{S}(2\vec{r}) \Gamma \gamma_3} \, , \quad\quad \psi \rightarrow e^{-\alpha\phi_\mathcal{S}(2\vec{r}) \Gamma\gamma_3} \psi \, .
    \label{DRot}
\end{equation}
This is the first order of the series of rotations (each with faster oscillations in space) which turn $\mathcal{S}_a$ to a field appearing only on the boundary of $\tri$s (see~\ref{fig:Fields}. Such a singular field on the boundary allows for the existence of flat bands when the magnetic field $B$ is changing sign by crossing the boundaries of $\tri$s, averaging to zero across the space~\cite{ZeroFlux}. The theory after the above rotation becomes,
\begin{equation}
    I = \int  \mathcal{D} \bar{\psi}\psi \bar{J}_\alpha \exp\left\{\int  d^2x   \bar{\psi} i \gamma^a \left( \partial_a + i \alpha\partial_a \phi_{\mathcal{A}} \cos (2\alpha\phi_\mathcal{S}(2\vec{r})) \Gamma \gamma_5 + \bar{\mathcal{S}}^1_a i\gamma_3 +\alpha\partial_a \phi_\mathcal{A} \sin(2\alpha\phi_\mathcal{S}(2\vec{r})) \Gamma i\gamma_3 \gamma_5 \right)\psi \right\}  \, .
\end{equation}
with $\bar{\mathcal{S}}^1$ being the new spin field which is one order closer to being constant across $\tri$ and singular on its boundaries. The new magnetic field giving rise to the magic angle is,
\begin{equation}
    \bar{B} \equiv \partial_x [\alpha\partial_y \phi_\mathcal{A}\cos(2\alpha\phi_\mathcal{S}(2\vec{r})] - \partial_y [\alpha\partial_x \phi_\mathcal{A}\cos(2\alpha\phi_\mathcal{S}(2\vec{r}))]
\end{equation}
depicted in Fig.~\ref{fig:Fields}. The approximation to the first magic angle with respect to $\bar{B}$ takes us step further in accuracy giving $\alpha \approx 0.589$.

\bibliographystyle{apsrev4-2}
\bibliography{supplement}